\documentclass[twocolumn,
aps,
prx,
superscriptaddress,
showpacs,
amsmath,
amssymb,
floatfix,
longbibliography,
eprint]
{revtex4-1}

\usepackage{appendix}
\usepackage{xcolor}
\definecolor{darkblue}{RGB}{0,0,150}
\definecolor{nightblue}{RGB}{0,0,100}

\usepackage{graphicx,mathtools,bm,bbm}
\usepackage{MnSymbol}
\usepackage[
colorlinks,
citecolor=darkblue,
linkcolor=darkblue,
urlcolor=nightblue]{hyperref}

\usepackage[english]{babel}
\usepackage[babel,kerning=true,spacing=true]{microtype}
\usepackage[utf8]{inputenc}

\usepackage{soul}

\newcommand{\ket}[1]{{\left|#1\right\rangle}}

\newcommand{\bra}[1]{{\left\langle #1\right|}}

\newcommand{\mytitle}{Gapping fragile topological bands by interactions}

\begin{document}
    
\title{\mytitle}
    
\author{Ari M.~Turner}
\thanks{All the authors have contributed equally to this work.}
\affiliation{Physics Department, Technion, Haifa 320003, Israel}
\author{Erez Berg}
\thanks{All the authors have contributed equally to this work.}
\affiliation{Department of Condensed Matter Physics, Weizmann Institute of Science Rehovot 7610001, Israel}
\author{Ady Stern}
\thanks{All the authors have contributed equally to this work.}
\affiliation{Department of Condensed Matter Physics, Weizmann Institute of Science Rehovot 7610001, Israel}

\begin{abstract}
We consider the stability of fragile topological bands protected by space-time inversion symmetry in the presence of strong electron-electron interactions. At the single-particle level, the topological nature of the bands prevents the opening of a gap between them. In contrast, we show that when the fragile bands are half filled, interactions can open a gap in the many-body spectrum without breaking any symmetry or mixing degrees of freedom from  remote bands. Furthermore, the resulting ground state is not topologically ordered. Thus, a fragile topological band structure does not present an obstruction to forming a ``featureless insulator'' ground state. Our construction relies on the formation of fermionic bound states of two electrons and one hole, known as ``trions''. The trions form a band whose coupling to the electronic band enables the gap opening. This result may be relevant to the gapped state indicated by recent experiments in magic angle twisted bilayer graphene at charge neutrality. 
\end{abstract}

\maketitle

\emph{Introduction.--} Topology has been recognized as a key player in condensed matter physics over the last few decades. 
%Given a certain set of symmetries, the electronic structure of a given material can fall into one of several ``topologically distinct'' classes, that cannot be continuously deformed into one another without the closing of an energy gap or the breaking of the symmetry. This topological structure can lead to a host of extremely robust physical phenomena, such as the quantization of the Hall conductivitiy and the existence of metallic states that are protected from gapping and localization as long as the symmetry is maintained. 
Much work has been devoted to classifying the topological properties of single-particle band structures~\cite{Schnyder2008,kitaev2009periodic,Hasan2010,Qi2011,Fu2011,Kruthoff2017,bradlyn2017topological,po2017symmetry}. However, it is always important to ask whether the topological structure of the single-particle states survives in the presence of strong electronic correlations~\cite{Fidkowski2010,Fidkowski2011,Turner2011,Fidkowski2013,kimchi2013featureless,Morimoto2015,Queiroz2016,Lee2018}. 
%In some examples, interactions fundamentally change the topological classification, 
%allowing to adiabatically connect phases that are topologically distinct without interactions, or to gap out zero-energy states whose gaplessness is protected in the non-interacting case~\cite{Fidkowski2010,Fidkowski2011,Turner2011,Fidkowski2013,kimchi2013featureless,Morimoto2015,Queiroz2016,Lee2018}.

In this work, we examine the stability of ``fragile topological bands''~\cite{Po2018Origin,Zou2018Band,Song2019,Po2019} in the presence of electron-electron interactions. A set of bands possesses fragile topology if {the subspace of their states} cannot be spanned by a complete basis of localized Wannier functions that preserves all the symmetries of the system, while a localized Wannier description exists upon adding topologically trivial bands \cite{Po2018,Ahn2019,Bouhon2019,Bradlyn2019}. Fragile topology arises in the nearly-flat bands in magic angle twisted bilayer graphene (MATBG)~\cite{
DosSantos2007,bistritzer2011moire,cao2018correlated,cao2018unconventional}.
%has attracted much attention recently~\cite{Po2018Origin,Zou2018Band,Song2019,Po2019}, since it arises in the nearly-flat bands in magic angle twisted bilayer graphene (MATBG)~\cite{DosSantos2007,bistritzer2011moire,cao2018correlated,cao2018unconventional}. 

Physically, fragile topology implies that without interactions, the set of bands cannot be separated {from one another} by an energy gap in the entire Brillouin zone (BZ) unless a symmetry is broken. Here, we find that in the presence of interactions, the system can become gapped even in the absence of any symmetry breaking or mixing with remote bands. Unlike interaction-induced gapping that arises elsewhere~\cite{Wang2013,bonderson2013time,Burnell2014,Metlitski2015,Mross2016}, here the gapped ground state is not required to have topological order. 

We provide a proof of principle for such a ``featureless insulator'' by introducing an interaction that uses states from the fragile bands to create a band of charged excitations composed of two electrons and one hole, and then coupling these excitations to the electronic band to open a gap in the many-body spectrum. 
The gap opens when the electron density in the system corresponds to an integer number of electrons per site, as required by the generalized Lieb-Schulz-Mattis theorem~\cite{lieb1961two,Oshikawa2000,Hastings2004}. Our mechanism is illustrated in two models that possess fragile topological bands protected by space-time inversion symmetry $C_2 \mathcal{T}_+$. The first model is defined on a square lattice and the second on a honeycomb lattice. The second model has the same symmetries and Wannier obstructions as MATBG~\cite{Zou2018Band}. 
\begin{figure*}[t]
\centering
\includegraphics[width=0.95\textwidth]{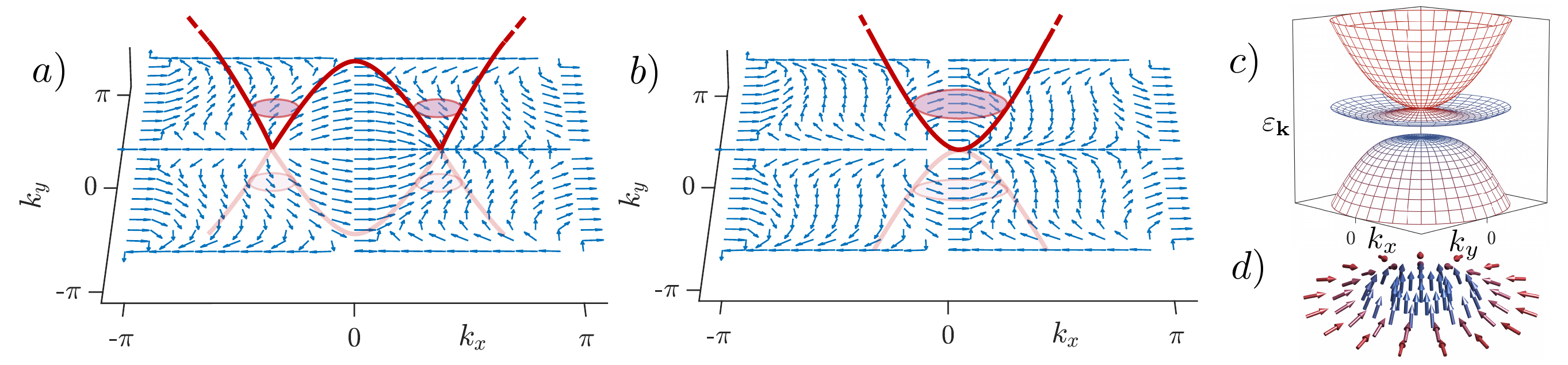}
\caption{(a,b) Winding of the vector  $(h_1(\bm{k}),h_2(\bm{k}))$ of the effective low-energy Hamiltonian [Eq.~\eqref{eq:Heff}] around the Dirac points.  %$H_{\rm{CB}}+\delta H$ [Eqs.~(\ref{squarelattice},\ref{squarelattice2})]. The arrows show the direction of the vector $(h_1(\bm{k}),h_2(\bm{k}))$ of the effective low-energy Hamiltonian [Eq.~\eqref{eq:Heff}]. 
(a) For $\Delta_x>0$ and $\Delta_{xz}=0$ [Eq.~\eqref{squarelattice2}], there are two Dirac points, both with winding number $+1$. (b) For $\Delta_x = -\Delta_{xz}$, the two Dirac points merge into a single quadratic band touching at $\bm{k}=0$ with a winding number of $+2$. (c) Spectrum of the three-band Hamiltonian~\eqref{eq:threebands}. 
When a third band is added, the quadratic band touching can be gapped, separating the lowest band from the other two. In our construction, the additional band consists of trion bound states. (d) Bloch wavefunction of the lowest band in panel (c). The wavefunction can be chosen to be real. The in-plane components represent the amplitudes of the two original electronic bands, and the out of plane component is the amplitude of the third (trion) band.}
\label{fig1}
\end{figure*}

{\it Model for fragile topological bands.--} The first model starts from two pairs of $C_2$--symmetric, spinless Chern bands of opposite Chern numbers~\footnote{We thank B. A. Bernevig, R. Queiroz, and Z. Song for explaining this model to us.}.   
The model has four orbitals on each site of a square lattice. 
The Hamiltonian is
\begin{equation}
    H_{\rm{CB}}({\bm k})=d_z({\bm k})\sigma_z + d_x({\bm k})\sigma_x\tau_z
    +d_y({\bm k})\sigma_y.
    \label{squarelattice}    
\end{equation}
Here, the Pauli matrices $\sigma,\tau$ act on the four-dimensional Hilbert space of each site, $d_z = m + t_1 \sin^2(k_x/2) + t_1\sin^2(k_y/2)$, $d_x = t_2 \sin(k_x)$, $d_y = t_2\sin(k_y)$, and $\bm{k}$ is measured in units of the inverse lattice spacing. For every $\bm k$, each band of $H_{\rm{CB}}$ contains two degenerate states, labeled by $\tau_z=\pm 1$. For positive $t_1$ and $-t_1<m<0$, the Chern numbers of the bands are $C=\pm\tau_z$. 
%, where the $\pm$ signs refer to the lower (upper) band. 

The Hamiltonian (\ref{squarelattice}) possesses  an anti-unitary symmetry $\mathcal{T}_+=K\tau_x \times (\bm{k}\rightarrow -\bm{k})$ (where $K$ is complex conjugation, and $\mathcal{T}_+^2=1$), as well as a unitary inversion symmetry $C_2=\sigma_z\times(\bm{k}\rightarrow -\bm{k})$, where $(\bm{k}\rightarrow -\bm{k})$ signifies a change in the sign of the momentum, and a mirror symmetry $M_y=\tau_x\times (k_x\rightarrow - k_x)$. 
%Additionally, it possesses an anti-unitary symmetry $\mathcal{T}_-=K\tau_y \times (\bm{k}\rightarrow -\bm{k})$ (with $\mathcal{T}_-^2=-1$). 
%, which does not play an important role in our analysis, and will be broken shortly. 

The crucial symmetry is the product $C_2{\mathcal T}_+$, which leaves the momentum unchanged, and in momentum space may be written as $C_2 \mathcal{T}_+ = K\tau_x\sigma_z$. 
We focus on the pair of low-energy bands, denoting their two degenerate Bloch states by $\vert \psi_\pm(\bm{k}) \rangle$ ($\pm$ denotes the eigenvalue of $\tau_z$). These states may be chosen such that $\vert \psi_- (\bm{k})\rangle= C_2 \mathcal{T}_+ \vert \psi_+(\bm{k}) \rangle$. 
Explicitly, we choose $\ket{\psi_+(\bm{k})} = \left(\cos\frac{\theta_{\bm k}}{2}, \sin\frac{\theta_{\bm k}}{2} e^{i\phi_{\bm{k}}} \right)^T\otimes \ket{\tau^z=1}$, where $\cos\theta_{\bm k} = -d_z({\bm k})/|\bm{d}(\bm{k})|$ and $e^{i\phi_{\bm k}} = -(d_x + id_y)/|d_x + id_y|$.

We add to (\ref{squarelattice}) two terms,
%which break $\mathcal{T}_-$ but preserve all other symmetries,
\begin{equation}
    \delta H=\Delta_x\tau_x+\Delta_{xz}\tau_x\sigma_z.
    \label{squarelattice2}
\end{equation}
%The combined Hamiltonian $H_{\rm{CB}}+\delta H$ may be diagonalized (see Appendix), but 
The essential features of $H_{\rm{CB}}+\delta H$ can be understood at the level of first-order degenerate perturbation theory in $\delta H$.
We define Pauli matrices $\eta_{x,y,z}$ that act in the subspace spanned by $\ket{\psi_{\pm}}$, such that $\eta^z \ket{\psi_{\pm}} = \pm \ket{\psi_{\pm}}$ and $\eta^x\ket{\psi_{\pm}}=\ket{\psi_{\mp}}$. The symmetry $C_2\mathcal{T}_+$ acts in this subspace as $\eta_xK$. 
Hence, the effective Hamiltonian projected to the two lowest-energy bands is% of the form 
\begin{equation}
    H_{\rm{eff}}(\bm{k}) = h_0(\bm{k}) + h_1(\bm{k})\eta_x + h_2(\bm{k})\eta_y, 
    \label{eq:Heff}
\end{equation}
where $h_0 = \bra{\psi_+} (H_{\rm{CB}} + \delta H) \ket{\psi_+}$, $h_1 - i h_2 = \bra{\psi_+} \delta H \ket{\psi_-}$. 
A third term proportional to $\eta_z$ is ruled out by $C_2\mathcal{T}_+$.

The eigenstates of Eq.~\eqref{eq:Heff} are determined by the direction of the planar vector field $(h_1(\bm{k}),h_2(\bm{k}))$. The two bands touch when $h_1(\bm{k}) = h_2(\bm{k}) = 0$. Each band touching can be  characterized by the winding of the phase $\alpha_{\bm k} \equiv \arg(h_1 + i h_2)$ as $\bm{k}$ is taken around the band touching point. A map of $\alpha_{\bm{k}}$ vs. $\bm{k}$ is shown in Fig.~\ref{fig1}a for $\Delta_x > 0$ and   $\Delta_{xz}=0$, for which there are two Dirac points in the BZ. The two Dirac points have the same winding number, $+1$. Consequently, even when the Dirac points coincide at $\bm{k}=0$ (for $\Delta_x = -\Delta_{xz}$), they do not annihilate each other to form a gap; rather, they form a quadratic band touching with a winding number of $+2$ (see Fig.~\ref{fig1}b). 
{The fact that the total winding number of all band touchings is non-zero indicates that the two bands possess fragile topology~\cite{Po2018}.}
%Note that this implies two related properties. First, the subspace of the two bands cannot be spanned by localized Wannier functions that preserve the $C_2 \mathcal{T}_+$ symmetry. Second, the two bands cannot be separated from one another by a gap unless the $C_2 \mathcal{T}_+$ symmetry is broken.} 

A nonzero net winding number sounds impossible at first,
since the total winding number around all the singular points of $\alpha_{\bm{k}}$ in the BZ must vanish due to its periodicity. However, there are additional singular points that are not associated with band touchings. Such points occur at momenta for which the basis wave functions $\ket{\psi_{\pm}(\bm{k})}$ are singular. The existence of such singularities %singular points in $\ket{\psi_{\pm}(\bm{k})}$ 
is a consequence of $\ket{\psi_{\pm}(\bm{k})}$ spanning Chern bands.
%necessitated by the fact that these wave functions span two Chern bands.  
For our choice of gauge, $\ket{\psi_{\pm}}$ are singular at $\bm{k}=(\pi,0)$, $(0,\pi)$, $(\pi,\pi)$, and
%, and 
$\alpha_{\bm{k}}$ winds around these points (see Fig.~\ref{fig1}a,b). 

{\it Interaction-induced insulating states.--} Physically, the fragile topological character of the lowest-energy bands implies that as long as the chemical potential lies within these bands, the system must be a (semi-)metal. Interactions between electrons can open a gap at a filling of one electron per unit cell, either by spontaneously breaking the $C_2 \mathcal{T}_+$~\cite{Xie2020} or translational~\cite{Kang2020} symmetry, or by  mixing remote bands~\cite{Xie2020}.

Here, we ask whether a gap can form even without breaking any symmetry or involving degrees of freedom beyond those of the two fragile bands. Furthermore, can such an insulating state have a unique ground state on a torus (a ``featureless insulator''), or does it have to possess topological order? 

Explicitly, we consider the Hamiltonian (\ref{eq:Heff}) for the two lowest bands at a filling of one electron per site, and add arbitrary finite-range interactions. We ask whether there is an insulating ground state that preserves $C_2 \mathcal{T}_+$, translation, and charge conservation symmetries. We allow modifications of the band structure that preserves these symmetries. 
%and does not close the single-particle energy gap between the two lower and two higher bands. 
Using this freedom, we shift the two Dirac points to the $\Gamma$-point by choosing $\Delta_{xz}=-\Delta_x$. Furthermore, $h_0({\bm k})$ is chosen such that there is no $\bm{k}\ne 0$ whose energy  $\varepsilon_{\bm{k}}$ equals $\varepsilon_{\bm{k}=0}$.
%is degenerate with the energy at the $\Gamma$ point. 

Our affirmative answer to the question above relies on the following strategy. We note that the band touching at $\Gamma$ is protected only as long as there are no additional bands to mix with. Consequently, a gap may open if the interactions effectively generate another band with low-energy states near $\Gamma$, consisting of degrees of freedom from other parts of the BZ. Such a band can be formed when the interactions produce ``trions''~\cite{Kheng1993,Finkelstein1995,Huard2000}, namely bound states of two electrons and a hole whose momenta add up to the $\Gamma$ point. Trion bound states, known to occur in certain semiconductors, are fermionic, charge-$e$ states. 
%, which can appear at low energy and momentum. 
By hybridizing the trions with electronic states near $\bm{k}=0$, we show that a gap in the spectrum can open without breaking any symmetry and without affecting the structure of the filled band away from the $\Gamma$ point.

{\it Opening a gap by hybridization with an additional band.--} We now show that the presence of a third trivial band near $\Gamma$ allows for the desired energy gap.  
The third band is taken to have a parabolic dispersion, $\varepsilon_{\rm{t}}(\bm{k}) = -\varepsilon_{\rm{t},0} + \frac{k^2}{2m_{\rm{t}}}$, where
$\varepsilon_{t,0}>0$. The subscript `t' stands for the third band, but also  anticipates the trions. 
We perform a unitary transformation on the two fragile bands, $U = e^{i \pi \eta_y/4} e^{i \pi \eta_x/4}$. Then, $C_2 \mathcal{T}$ changes to $U (\eta_x K) U^\dagger = i K$, and the Hamiltonian is real. Near $\bm{k}=0$, it is
\begin{equation}
    H_{\rm{3-bands}}(\bm{k})=\left (
    \begin{array}{ccc}
\frac{k_y^2-k_x^2}{2 m} &  -\frac{k_x k_y}{m}         & g_1(\bm{k})  \\
-\frac{k_x k_y}{m}   &    \frac{k_x^2-k_y^2}{2 m}     & g_2(\bm{k}) \\
g_1(\bm{k}) & g_2(\bm{k}) & \frac{k^2}{2m_{\rm{t}}}-\varepsilon_{\rm{t},0}
    \end{array} \right ).
    \label{eq:threebands}
\end{equation}
Here, the upper left $2\times 2$ block gives the uncoupled quadratic band touching at $\bm{k}=0$, with $m \propto 1/\Delta_x$. Choosing $(g_{1}, g_2) \propto (k_x, k_y)$ hybridizes the third band with the lowest of the two original bands, separating them by a gap~\cite{SI}. The resulting spectrum is shown in Fig.~\ref{fig1}c. 
%We note that the precise form of $(g_{1}, g_{2})$ is not important, since once a 
%gap is formed, it is stable to perturbations. 
This mechanism of opening a gap is related to the ones discussed in Refs.~\cite{wu2019non,Ahn2019,bouhon2020non,Kang2020}.

The eigenstates of (\ref{eq:threebands}) may be chosen to be real. 
%Since the Hamiltonian (\ref{eq:threebands}) is real, the eigenstates may be chosen to be real, in which case they 
As such, they correspond to unit vectors in three dimensions. For $g_1=g_2=0$, the vectors that correspond to each of the two original bands lie in the $x-y$ plane, and form a vortex structure around $\Gamma$. The limit ${\bm k}\rightarrow 0$ is then singular. This enforces a band touching between the two original bands.  
%For a real and single-valued choice of the wave functions in the entire Brilloiun zone to be possible, there must be more singular points around which the 2D vector that describes the wave function winds, such that the total vorticity vanishes. 
The hybridization with the third band allows the unit vector that corresponds to the wave function to acquire a $z$--component, and turns the vortex into a meron (Fig.~\ref{fig1}d). Importantly, the hybridization changes the lowest-energy band only at small momenta, and leaves both the energies and the wave functions largely unaffected far from $\Gamma$~\footnote{Such a gapping mechanism is impossible near a single Dirac point of the original bands; in that case, the lowest energy band has a Berry phase of $\pi$ going around the Dirac point, which enforces a band touching between that band and the bands above it.}.

{\it Formation of a trion band.--} 
Returning to our original problem with two isolated bands 
[Eq.~\eqref{eq:Heff}], we now show that interactions can drive the system into an insulating phase without breaking any symmetry. This is done through the formation of a low-momentum band of trions. 
%a bound state of two electrons to one hole to create a fermionic excitation called ``trion" (which is known to occur in semiconductors). 
The trions provide the additional fermionic degree of freedom near $\Gamma$ needed to gap out the quadratic band touching of the two original bands. {Importantly, within our construction, the trions and the electrons near the quadratic band touching are independent degrees of freedom, since the trions are made of states whose momenta are far from the quadratic band touching.} 

%We now proceed to formulate an interaction that binds two electrons to one hole to create a fermionic excitation which we call a "trion", following similar excitations in semi-conductors. 

\begin{figure}[t]
\centering
\includegraphics[width=.48\textwidth]{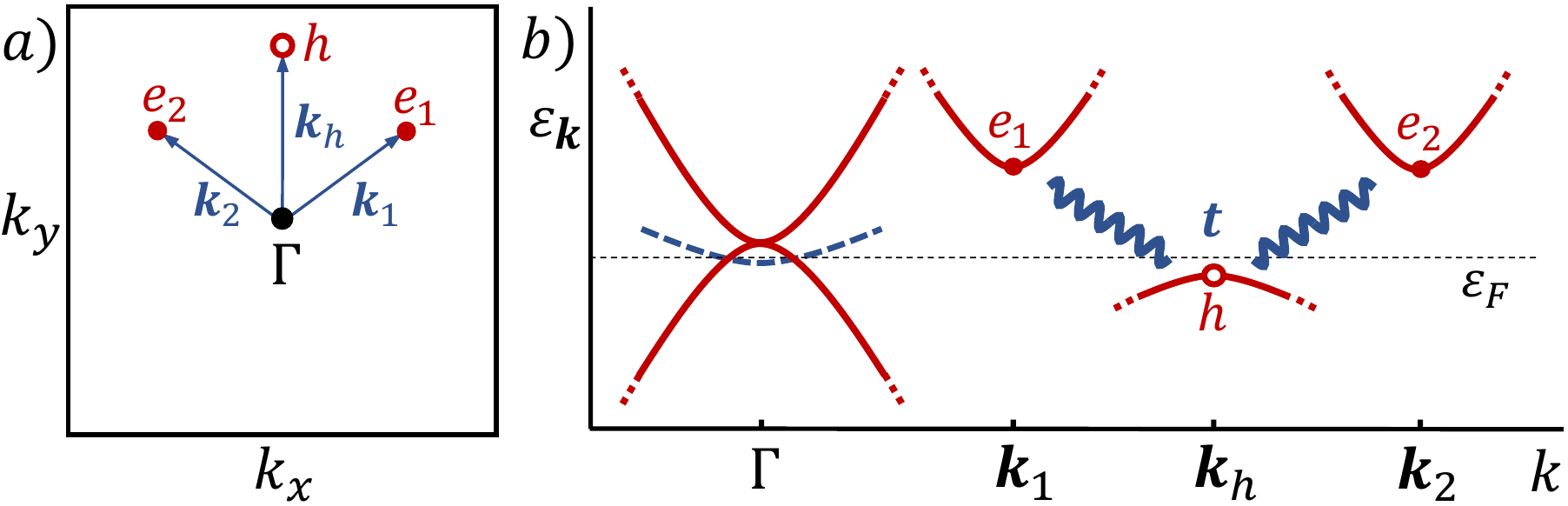}
\caption{(a) To facilitate the formation of trion bound states near $\bm{k}=0$, we first deform the single-particle dispersion to contain two electron valleys centered at $\bm{k}_{1,2}$, and a hole valley at $\bm{k}_{\rm{h}} = \bm{k}_1 + \bm{k}_2$. (b) Schematic cut of $\varepsilon_{\bm{k}}$ on a trajectory $\Gamma\rightarrow\bm{k}_{1}\rightarrow \bm{k}_{\rm{h}}\rightarrow \bm{k}_2$
. 
%passing through $\Gamma$, $\bm{k}_{1,2}$ and $\bm{k}_h$. 
The trions form due to interactions between the two electron valleys and the hole valley. The minimum of the trion dispersion is at $\bm{k}=0$ (dashed blue line).}
\label{fig2}
\end{figure}

To create the trions, imagine deforming the single-particle dispersion to include two local minima in the conduction band at $\bm{k}_{1,2}$ away from $\bm{k}=0$, and a local maximum in the valence band at $\bm{k}_{\rm{h}}=\bm{k}_{1}+\bm{k}_{2}$ (see Fig.~\ref{fig2}a). The dispersions close to $\bm{k}_{1},\bm{k}_{2},\bm{k}_{\rm{h}}$ are quadratic, and are taken to be $\varepsilon_{i,\bm{k}} = \frac{(\bm{k} - \bm{k}_{i})^2}{2m_{\rm{e}}} + \varepsilon_{\rm{e},0}$ (where $i=1,2$) and $\varepsilon_{\mathrm{h},\bm{k}} = -\frac{(\bm{k} - \bm{k}_{\rm{h}})^2}{2m_{\rm{h}}} - \varepsilon_{\rm{h},0}$. Here, $m_{\rm{e,h}}$ are the effective masses of the electron and hole valleys, and $\varepsilon_{\rm{e},0},\varepsilon_{\rm{h},0}>0$ are their energy offsets relative to the Fermi energy (Fig.~\ref{fig2}b).

We choose the interaction 
%between electrons and holes 
in the vicinity of the three valleys to couple the electrons only to the holes, $V_{\rm{eff}} = - u_0 \sum_{i=1,2} \int d^2r \, \psi^\dagger_{{\mathrm e},i} \psi^\dagger_{\rm{h}} \psi^{\vphantom{\dagger}}_{\rm{h}} \psi^{\vphantom{\dagger}}_{{\mathrm e},i}$, where $\psi_{{\mathrm e},i}^\dagger(\bm{r})$ creates an electron in a wavepacket centered at $\bm{r}$ whose momenta are localized near $\bm{k}_{i}$, $\psi_{\rm{h}}^\dagger(\bm{r})$ similarly creates a hole in the valence band near $\bm{k}_{\rm{h}}$, and $u_0>0$ is a coupling constant. 
%We will comment on the effects of interactions between electrons in the two valleys at $\bm{k}_{i=1,2}$ later. 

This interaction creates bound states, excitons and trions. The excitons are composed of one electron and one hole, carrying a non-zero momentum near $\bm{k}_{i} + \bm{k}_{\rm{h}}$, with binding energy $\varepsilon_{\rm{b}}$~\footnote{In $d=2$, any attractive interaction between electrons and holes yields an exciton bound state.}. The bottom of the excitation band is at energy $\varepsilon_{\rm{ex}} = \varepsilon_{\rm{e},0} + \varepsilon_{\rm{h},0} - \varepsilon_{\rm{b}}$; to avoid the excitons from condensing and breaking translational symmetry, we require $\varepsilon_{\rm{ex}}>0$. 
The trion is composed of two electrons and one hole, and carries a momentum near zero.  
Setting, for simplicity, $m_{\rm{h}}\gg m_{\rm{e}}$, the bottom of the trion band is at $-\varepsilon_{\rm{t},0} = 2\varepsilon_{\rm{e},0} + \varepsilon_{\rm{h},0} - 2\varepsilon_{\rm{b}}$. 
The combination of the two requirements $\varepsilon_{\rm{ex}}>0$ and $\varepsilon_{\rm{t},0}<0$ gives
$\varepsilon_{\rm{e},0}+\frac{1}{2}\varepsilon_{\rm{h},0}<\varepsilon_{\rm{b}}<\varepsilon_{\rm{h},0} + \varepsilon_{\rm{e},0}$. 
Close to $\bm{k}=0$, the trion energy is $\varepsilon_{\rm{t}}(\bm{k}) = -\varepsilon_{\rm{t},0} + \frac{k^2}{2m_{\rm{t}}}$. Due to the Galilean invariance of our model near the electron and hole valleys, $m_{\rm{t}} = 2m_{\rm{e}} + m_{\rm{h}}$. At large enough momenta, the trion band merges into the two-electron, one-hole continuum, and ceases to be a well-defined bound state. Since $\varepsilon_{\rm{t},0}>0$, the many-body ground state includes a non-zero density of trions in a pocket centered at $\bm{k}=0$. 

The final step in our construction couples the trions to electrons in the lower original band near $\bm{k}=0$, in the way described in Eq.~\eqref{eq:threebands}. The trion creation operator in real space is
\begin{eqnarray}
    t^\dagger(\bm{r}_{\rm{t}}) &=& \int d^2{r_{\rm{e},1}}d^2{r_{\rm{e},2}}\, F^\star(\bm{ r}_{\rm{e},1}-\bm{r}_{\rm{t}},{\bm{r}_{\rm{e},2}-\bm{r}_{\rm{t}}})\nonumber \\ &\times& \psi^\dagger_{\rm{h}}({\bm{r}_{\rm{t}}})\psi^\dagger_{\rm{e},1}(\bm{ r}_{\rm{e},1}-\bm{r}_{\rm{t}})\psi^\dagger_{\rm{e},2}({\bm{r}_{\rm{e},2}-\bm{r}_{\rm{t}}}),
    \label{eq:trion_wf}
\end{eqnarray}
where the trion wavefunction $F(\bm{r},\bm{r}')$ decays at long distances. 
We introduce an interaction $V_{\rm{t}-\Gamma}$ that hybridizes the trions with the electrons near the $\Gamma$ point:
\begin{equation}
V_{\rm{t}-\Gamma} = u \sum_{j=1,2}\int d^2r\,t^\dagger(\bm{r})  \frac{1}{i}\partial^{\vphantom{\dagger}}_{r_j} \psi^{\vphantom{\dagger}}_j(\bm{r}) + \rm{h.c.}
\label{eq:VtG}
\end{equation}
with a coupling constant $u$. The operators $\psi^{\dagger}_{j={1,2}}(\bm{r})$ create electrons near $\Gamma$ (the subscript labels the two eigenstates of $\eta_z$), and $r_{j=1,2}$ are the two spatial components of $\bm{r}$. 
This interaction has the same effect as the $g_{1,2}$ terms in Eq.~\eqref{eq:threebands}. %Correspondingly, . The spatial derivatives are the Fourier transforms of $g_{1,2}(\bm{k})$. 
Note that $V_{\rm{t}-\Gamma}$ is a two-particle short-range interaction, as can be seen by inserting Eq.~\eqref{eq:trion_wf} into Eq.~\eqref{eq:VtG}. 

\begin{figure}[t]
\centering
\includegraphics[width=.35\textwidth]{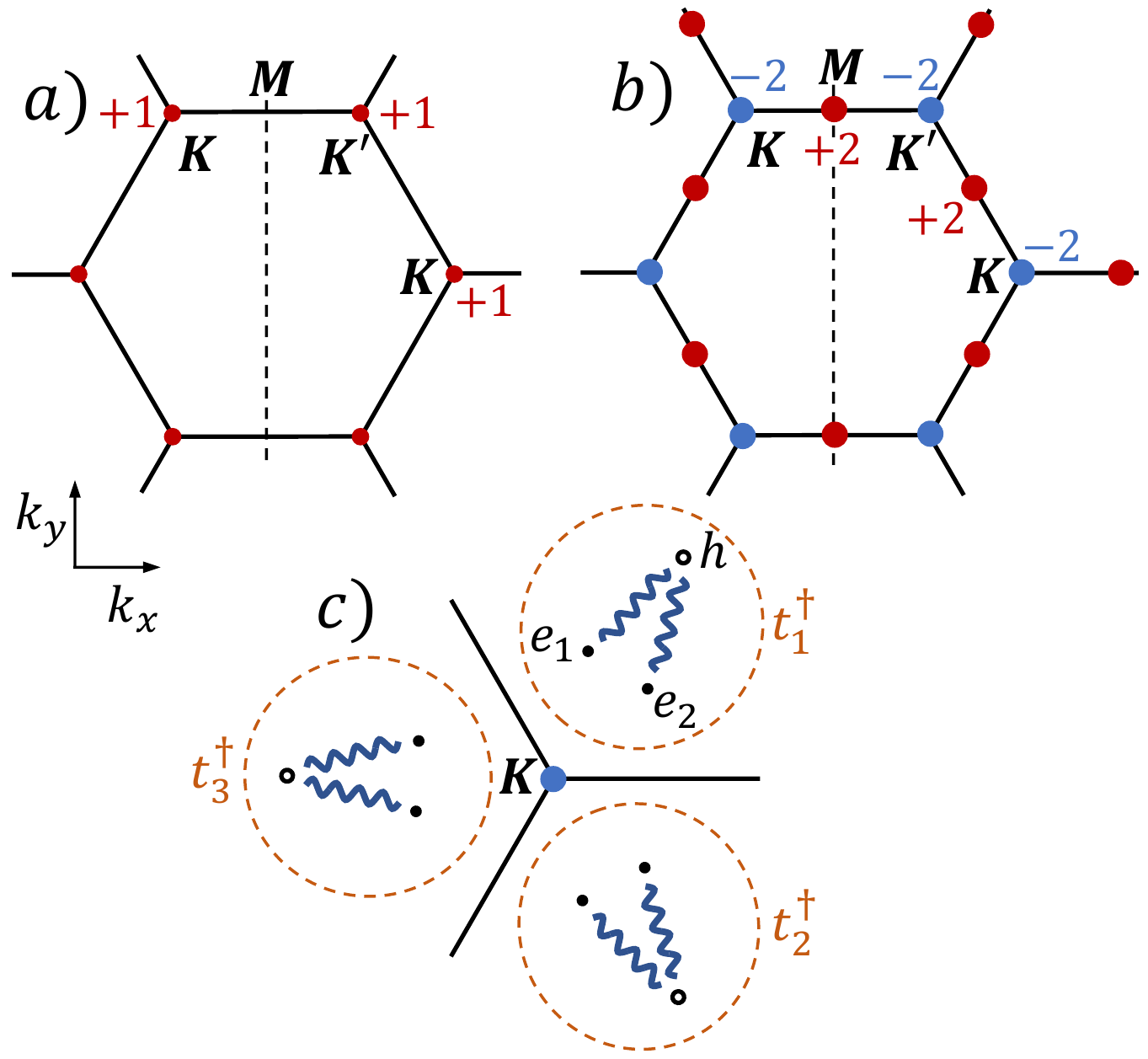}
\caption{Gapping fragile topological bands with $C_3$ symmetry. (a) BZ of the four-band model on the honeycomb lattice~\cite{SI}. The lower two bands touch at Dirac points (red) at $K$ and $K'$, whose winding numbers are $+1$. (b) The two fragile bands can be deformed into a configuration with quadratic touchings at $K$ and $K'$ whose winding numbers are $-2$ (blue), and quadratic touchings at the $M$ points with winding $+2$. 
%, such that the total winding number in the entire zone remains $+2$. 
(c) To gap out the quadratic touching at $K$, we form three trion bands $t_{1,2,3}$ in a $C_3$ symmetric fashion around $K$. The symmetric combination $\frac{1}{\sqrt{3}}(t_1 + t_2 + t_3)$ can hybridize with the lower band at $K$ without breaking $C_3$, opening a gap in the spectrum (see Fig.~\ref{fig1}c). The quadratic touching at $K'$ may be gapped in a similar way. The quadratic touching at $M$ can be gapped by introducing a pair of electron valleys and a hole valley in a mirror symmetric way, forming a trion band, and hybridizing it with the electrons at $M$.}
%The other quadratic band touchings can be similarly gapped by forming additional trion bands.}
\label{fig3}
\end{figure}

The interaction $V_{\rm{t}-\Gamma}$ achieves our goal: %exactly what we set out to do: 
it opens a gap in the many-body spectrum, leading to a featureless insulating ground state without involving degrees of freedom external to the two fragile topological bands. 

Several comments are in order. First, our treatment of the trions as essentially free fermions is justified in the limit where their size is much smaller than the spacing between them. 
{In this limit, the interaction energy between trions can be neglected, because it scales to zero faster than the gap~\cite{SI}.}
Second, the resulting ground state is not a Slater determinant in terms of the original electrons. However, the wavefunction deviates significantly from a Slater determinant only in the vicinity of $\Gamma,\bm{k}_{1},\bm{k}_{2},\bm{k}_{\rm{h}}$. The filled states near $\Gamma$ are depicted in Fig.~\ref{fig1}d, with the $z$ component proportional to the trion amplitude. Importantly, the spectrum does not contain any fractionalized excitations, and hence the system is not topologically ordered. Finally, our construction relies on merging the two Dirac points into a quadratic band touching. This is impossible when the Dirac points are pinned to a particular point in the BZ, e.g., due to $C_3$ symmetry (as in MATBG). This case is treated next.

\emph{Gapping in the presence of $C_3$ symmetry.--} To address this case, we use a $C_3$ symmetric model that gives a pair of fragile topological bands on a honeycomb lattice, described in Ref.~\cite{Zou2018Band} (see ~\cite{SI}). As before, the fragile topology is protected by $C_2\mathcal{T}_+$ and translational symmetries. Fig.~\ref{fig3}a shows the BZ, indicating the positions of the two Dirac points at $K$ and $K'$ and their winding numbers, which are both $+1$. Importantly, the Dirac points cannot be moved away from $K$ and $K'$ without breaking the $C_3$. 

%As we now explain, the pinning of the Dirac points does not prohibit their gapping by interactions. 
To demonstrate that this does does not prohibit gapping by interactions, we deform the single-particle Hamiltonian, turning $K,K'$ into quadratic touching points at the cost of adding three additional quadratic touchings half-way between $K$ and $K'$, at the $M$ points (Fig.~\ref{fig3}b). 
The deformation from Fig.~\ref{fig3}a to Fig.~\ref{fig3}b can be performed by first nucleating at $M$ two Dirac points with winding numbers $-1$ and two with winding numbers $+1$. The $-1$ Dirac points are pulled apart and moved towards the $K$ and $K'$ points, leaving at $M$ quadratic band touching with winding $+2$. The $K$ and $K'$ points absorb three Dirac points with winding $-1$ each, consistently with $C_3$ symmetry, turning into quadratic band touchings with winding numbers of $-2$~\cite{anim}. With all gap-closures being quadratic, the interaction-driven gapping mechanism described above can be applied, as explained in Fig.~\ref{fig2}. 

\emph{Discussion.--} We end with a brief discussion of the possible implications of our results for magic-angle twisted bilayer graphene. The narrow bands in this system have a fragile topological character of the type discussed here~\cite{Po2018,Zou2018Band}. This system has an additional spin and valley multiplicity. Our mechanism for opening a gap without symmetry breaking is, in principle, applicable for each of the spin and valley flavors separately~\footnote{This multiplicity allows for an additional gapping mechanism by spontaneously breaking the valley charge conservation~\cite{Bultnick2020}.}. 

Whether there is a gap at charge neutrality in MATBG is presently unresolved, with some experiments observing semi-metallic behavior~\cite{cao2018correlated} and others observing a gap~\cite{lu2019superconductors}. Our work implies that a gap may occur even in the absence of any broken symmetry. The mechanism we outlined here serves as a proof of principle. Given the strongly interacting nature of MATBG, it is possible that other, less fine-tuned  mechanisms may lead to the same gapped, symmetry-preserving state. {For instance, a variant of our mechanism can be used to construct a featureless insulating state on the honeycomb lattice at half filling \cite{SI}; the same phase can be constructed via a different mechanism that does not involve trions~\cite{kimchi2013featureless,Lee2018}}.

Our mechanism for a gap opening invokes the formation of an additional band of trions. Since the trion has the same quantum numbers as an electron, the trion band should be observable in angle-resolved photoemission or scanning tunneling microscopy experiments. 

To conclude, our work elucidates an aspect in which ``fragile topology'' lives up to its name: while the topological character prevents the bands from being separated by a gap at the single-particle level, no such fundamental obstruction exists once interactions are included. 

\acknowledgements{
%{\it Acknowledgements.--}
We thank B. Andrei Bernevig, Eslam Khalaf, Raquel Queiroz, Robert-Jan Slager, Zhida Song, and Oskar Vafek for useful discussions. E.B. and A.S. acknowledge support from the Israeli Science Foundation Quantum Science and Technology grant no. 2074/19, and from the
CRC 183 of the Deutsche Forschungsgemeinschaft (Project C02). 
This project has received funding from the European Research Council (ERC) under the European Union’s Horizon 2020 research and innovation programme [grant agreements No. 788715 and 817799, Projects LEGOTOP (AS) and HQMAT (EB)]. 
 A.M.T. acknowledges support from the Israeli Science Foundation grant no. 1939/18.}

%- Measureable consequences: additional lines in ARPES.

\bibliography{TBG_refs}

\clearpage

\appendix
\renewcommand{\thefigure}{S\arabic{figure}}
\renewcommand{\figurename}{Figure}
\setcounter{figure}{0}
\setcounter{equation}{0}
\renewcommand{\theequation}{S\arabic{equation}}

\section*{Supplementary Information}

\subsection{Colormap representation of the winding around the Dirac points}

\begin{figure}[t]
\centering
\includegraphics[width=.45\textwidth]{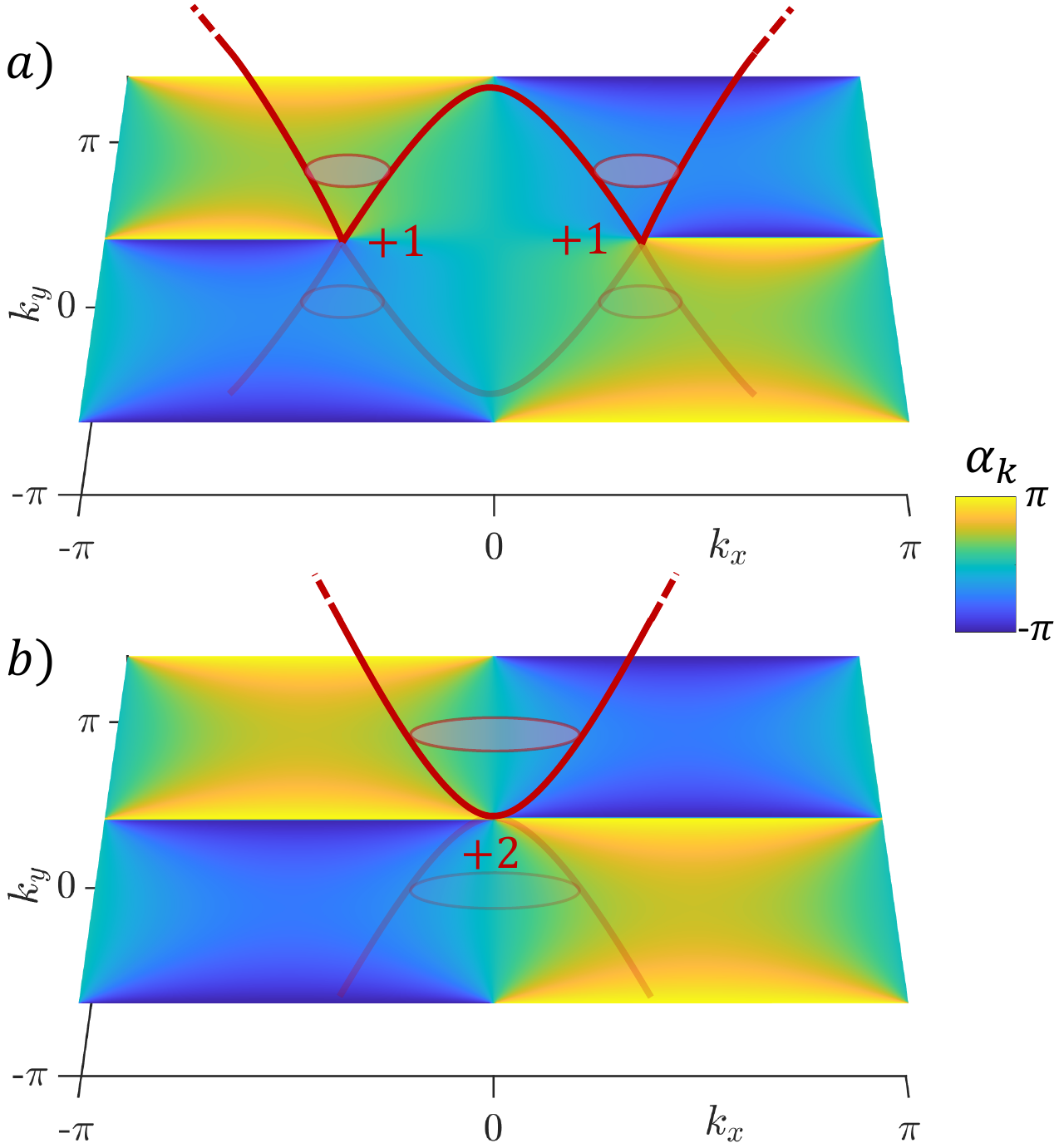}
\caption{The angle $\alpha_{\bm{k}}$ for the square lattice model as a function of $\bm{k}$, represented by color. The same angle is represented in Fig.~1 of the main text by arrows. (a) $\Delta_x>0$, $\Delta_{xz}=0$; in this case, there are two Dirac points, both with a winding of $+1$ (indicated in red). (b) $\Delta_x = -\Delta_{xz}$. In this case, the Dirac points merge into a single quadratic band touching at $\bm{k}=0$ with winding number $+2$.}
\label{figS1}
\end{figure}

For the reader's convenience, we display the angle $\alpha_{\bm{k}} \equiv \arg(h_1(\bm{k})+ih_1(\bm{k}))$ as a colormap in Fig.~\ref{figS1}. The color scale ranges from $-\pi$ to $\pi$. Note that the lines of discontinuities in $\alpha_{\bm{k}}$ by $2\pi$, that emanate from the band touchings, are not physical. The winding numbers of the different band touchings are shown in the figure. The angle $\alpha_{\bm{k}}$ depends on our choice of phase for the wavefunctions of the two fragile bands, but the relative winding numbers of the Dirac cones do not depend on this gauge choice. 

\subsection{Diagonalization of the three-band Hamiltonian}

Here, we diagonalize the three-band Hamiltonian, Eq.~(4) in the main text. We choose $g_1(\bm{k}) = \Delta\,k_x$, $g_2(\bm{k}) = \Delta\, k_y$, where $\Delta$ represents the strength of the hybridization between the additional third band and the original two bands. The diagonalization is simplified by exploiting the transformation law of the Hamiltonian under rotation. In particular, $H_{\rm{3-band}}$ satisfies
\begin{equation}
H_{\rm{3-band}}\left(R_{\phi}\bm{k}\right) = R^T_{\phi} H_{\rm{3-band}}(\bm{k}) R_{\phi},
\label{eq:rot}
\end{equation}
where $R_{\phi}$ is the following orthogonal matrix:
\begin{equation}
    R_{\phi}=\left (
    \begin{array}{ccc}
\cos(\phi)  &  \sin(\phi)         &  0  \\
-\sin(\phi) &  \cos(\phi)         &  0  \\
0           &     0              &  1
    \end{array} \right ).
    \label{eq:R}
\end{equation}
Thus, to find the spectrum, it is sufficient to consider a momentum $\bm{k}$ along the $x$ axis. The spectrum at a different values of $\bm{k}$ can then be found by rotating ${\bm{k}}$ onto the $x$ axis by Eq.~\eqref{eq:rot}, and using the fact that the spectrum is invariant under an orthogonal transformation. 

\begin{figure}[t]
\centering
\includegraphics[width=.45\textwidth]{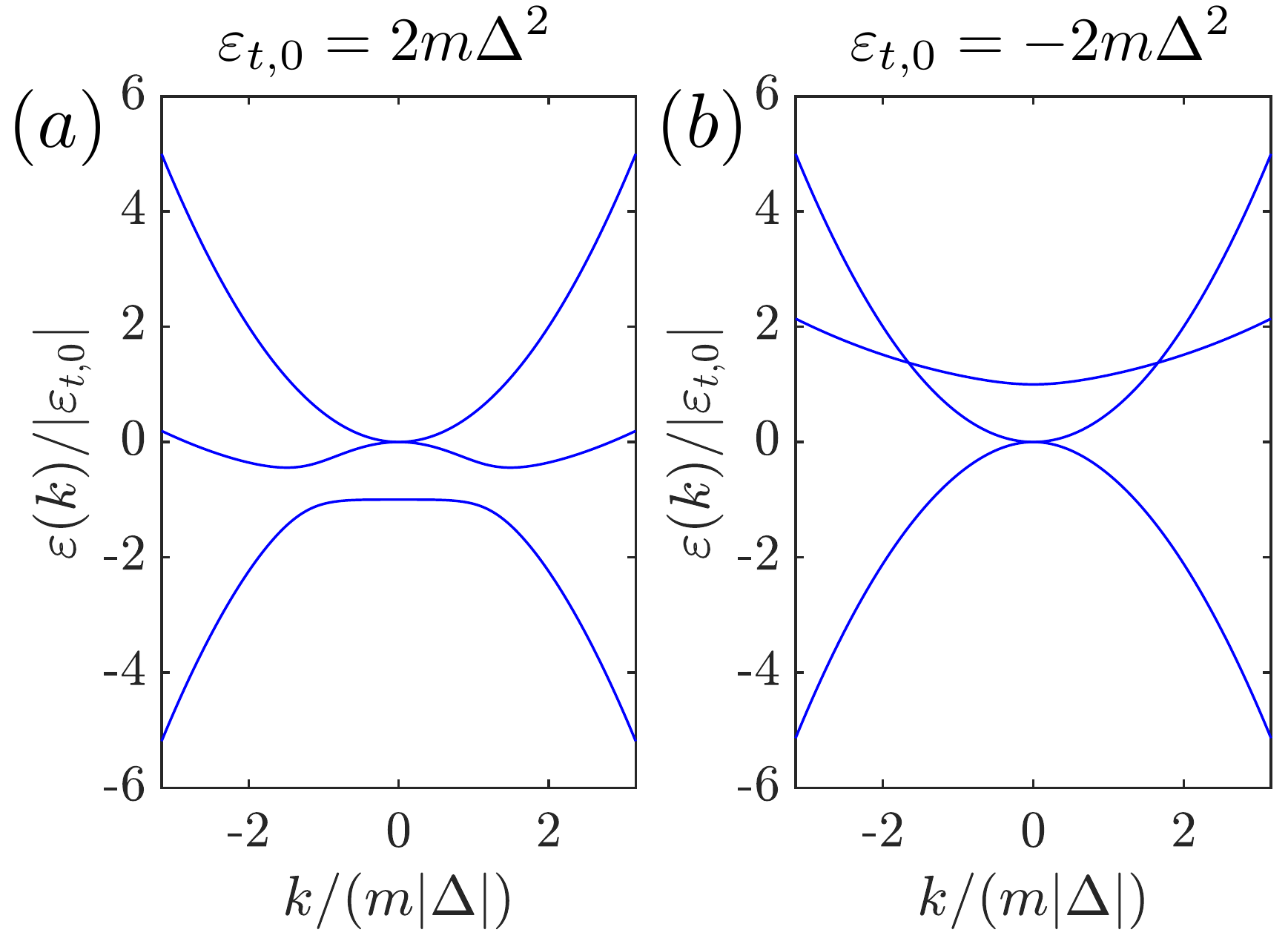}
\caption{Spectrum of $H_{\rm{3-band}}$, given in Eq.~(4) of the main text, with $g_1(\bm{k}) = \Delta\,k_x$ and $g_2(\bm{k}) = \Delta\,k_y$. With this choice, the spectrum is rotationally symmetric, depending only on $|\bm{k}|$. We have used $m_{\rm{t}} = 5m$. In panel (a), $\varepsilon_{\rm{t},0} = 2m \Delta^2$, whereas in (b),  $\varepsilon_{\rm{t},0} = -2m\Delta^2$.}
\label{figS2}
\end{figure}

From these considerations, we find that the three eigenvalues of $H_{\rm{3-band}}$ are given by
\begin{align}
\varepsilon_{1}(\bm{k}) & =\frac{k^{2}}{2m},\\    
\varepsilon_{2,3}(\bm{k}) &=\frac{1}{2}\left[\varepsilon_{\rm{t}}(k)-\frac{k^{2}}{2m}\right]\pm\sqrt{\Delta^{2}k^2+\frac{1}{4}\left[\varepsilon_{\rm{t}}(k)+\frac{k^{2}}{2m}\right]^{2}},\nonumber 
\end{align}
where $\varepsilon_{\rm{t}}(k)\equiv -\varepsilon_{\rm{t},0} + k^2/(2m)$. 
The spectrum is shown in Fig.~\ref{figS2}. As can be seen in the figure, the lowest band is separated from the other two bands by a gap if $\varepsilon_{\rm{t},0}>0$.

%It is worth noting that our choice of $g_{1,2}(\bm{k})$ makes the problem completely rotationally symmetric around $\bm{k}=0$, as is implied by Eq.~\eqref{eq:rot}. 
Our choice of $g_{1,2}(\bm{k})$, for which the problem is rotationally symmetric, is made mostly for convenience. It is not essential in order to demonstrate the possibility of separating the lowest of the three bands of $H_{\rm{3-band}}$ from the other two by a gap. We note, however, that in the $C_3$ symmetric case discussed in the main text, for the quadratic band touchings at $K$ and $K'$, the Hamiltonian near these points is required to be $C_3$ symmetric. The form we have chosen satisfies this automatically, as $C_3$ is a subgroup of the continuous rotational symmetry of $H_{\rm{3-band}}$. (One can check that within the $C_3$ symmetric model, described in Sec.~\ref{sec:C3} below, the states at the quadratic band touchings transform according to $R_{\phi}$ above when written in a real basis.) 

Interestingly, even after the third band hybridizes with the original two, the quadratic band touching at $\bm{k}=0$ persists. This is a general feature of our three-band problem. As long as $C_2 \mathcal{T}_+$ symmetry is preserved, and hence the Hamiltonian can be chosen to be real, three eigenstates of the Hamiltonian are specified by an orthogonal matrix that depends on $\bm{k}$. Since $\pi_1[O(3)] = \mathbb{Z}_2$, the eigenstates along a closed path that encircles the origin are characterized by a $\mathbb{Z}_2$ topological invariant. The presence of a quadratic band touching at the origin corresponds to this invariant being non-trivial. Therefore, a quadratic band touching between two of the three bands within the closed path cannot disappear unless $C_2 \mathcal{T}_+$ is broken or the band touching crosses the path. This feature generalizes to any number $N\ge 2$ of bands with $C_2 \mathcal{T}_+$ symmetry, such that the eigenstates can be chosen to be real. Note, however, that this does not preclude a many-body gapped state with an integer filling per site, since the band touching can occur either below or above the Fermi level. 

\subsection{Treatment of interactions between trions}

Here, we present a more systematic discussion
of the effective interacting Hamiltonian that describes the low-energy
degrees of freedom of the system. Crucially, we show that the interactions between the trion bound states can be neglected in the limit where the trions are dilute in the ground state, justifying the treatment of the trions as essentially non-interacting particles.  

The degrees of freedom near the chemical potential include the gapless electrons
near $\Gamma$ and the trions. Within our construction, there are no other low-energy degrees of freedom.
In particular, the electrons and holes near $\bm{k}_{1,e},$ $\bm{k}_{2,e}$
and $\bm{k}_{h}$, as well as any bosonic excitons formed by combining
an electron and a hole, remain gapped, as discussed in the manuscript. 

The low-energy effective Hamiltonian is written as
\begin{equation}
H_{{\rm eff}}=H_{\Gamma}+H_{t}+H_{\Gamma t}+H_{int}.\label{eq:Heff_SI}
\end{equation}
Here, $H_{\Gamma}$ is the bare Hamiltonian of electrons near $\Gamma$.
As in the main text, the basis is chosen such that the states near $\Gamma$ are spanned by $\vert u_{1,2}(\bm{k})\rangle$, which are smooth near $\Gamma$, and $C_2\mathcal{T}_+$ symmetry acts simply as
complex conjugation. 
The corresponding creation operators are $c_{\bm{k},1}^{\dagger}$, $c_{\bm{k},2}^{\dagger}$. In terms of these operators, $H_{\Gamma}$ is
written as
\begin{equation}
H_{\Gamma}=\sum_{|\bm{k}|\le k_{0}}c_{\bm{k}}^{\dagger}h^{\vphantom{\dagger}}_{\bm{k}}c^{\vphantom{\dagger}}_{\bm{k}},\label{eq:H_G}
\end{equation}
where $c_{\bm{k}}^{\dagger}=(c_{\bm{k},1}^{\dagger},c_{\bm{k},2}^{\dagger})$
and 
\begin{equation}
h_{\bm{k}}=\frac{1}{2m}\left(\begin{array}{cc}
k_{y}^{2}-k_{x}^{2} & k_{x}k_{y}\\
k_{x}k_{y} & k_{x}^{2}-k_{y}^{2}
\end{array}\right),
\end{equation}
forming the quadratic band touching. The momentum summation in (\ref{eq:H_G})
is restricted to a disk of radius $k_{0}$ around the $\Gamma$ point, such that
$k_{0}$ serves as the momentum cutoff of our low-energy theory. Without
the coupling to the trions, $H_{\Gamma}$ has a gapless spectrum. 

Next, we discuss the trions. A single trion excitation is created
by an operator $t_{\bm{k}}^{\dagger}$, which
is composed entirely of electron and hole states outside the disk
of radius $k_{0}$ of $\Gamma$, and hence can be treated as an independent
fermionic degree of freedom. Since the energy of the trions is
assumed to be minimal at $\bm{k}=0$, their effective Hamiltonian
can be written as
\begin{equation}
H_{t}=\sum_{|\bm{k}|\le k_{0}}\left(\frac{k^{2}}{2m_{t}}-\varepsilon_{t,0}\right)t_{\bm{k}}^{\dagger}t_{\bm{k}}.
\end{equation}
The parameters are chosen such that the characteristic momentum
of the trions, $k_{t}\equiv\sqrt{m_{t}\varepsilon_{t,0}}$, is much
smaller than $k_{0}$. To simplfy our discussion, we assume that $m\sim m_{t}$. 

The $H_{\Gamma t}$ term hybridizes the electrons and holes with the
trions, and is chosen to be
\begin{equation}
H_{\Gamma t}=\sum_{|\bm{k}|\le k_{0}}\Delta\, t_{\bm{k}}^{\dagger}\left(k_{x}c_{1\bm{k}}+k_{y}c_{2\bm{k}}\right)+h.c.,
\end{equation}
where $\Delta$ is real. We choose $\Delta\sim k_{t}/m_{t}\ll k_{0}/m_{t}$,
such the hybridization between electrons and trions outside of the
disk of radius $k_{0}$ is negligible, ensuring the consistency of
our low-energy treatment. In terms of the original electrons and holes,
$H_{\Gamma t}$ is a local two-particle interaction between an electron
near $\Gamma$ and electrons near $\bm{k}_{1,e},\bm{k}_{2,e},$ or
$\bm{k}_{h}$, as discussed below Eq. (6) of the manuscript. 

The trions necessarily interact with each other, as can be seen by considering the Fourier
transformed operator $t^{\dagger}(\bm{r})$ that creates a trion centered
at point $\bm{r}$ in real space (Eq. 5 in the manuscript). The trion
has a characteristic size $a_{t}\sim1/\delta k_{t}$, where $\delta k_{t}$
is the spread in momentum space around $\bm{k}_{1,e},\bm{k}_{2,e},\bm{k}_{h}$
of the states that participate in the trion excitation. $\delta k_{t}$
is determined by the trion binding energy (and hence by the strength of the
interaction between electrons and holes that binds them together to
form the trion). Two trions within $a_{t}$ of each other interact
due to their spatial overlap.

To account for this effect, we include an inter-trion effective interaction
\begin{equation}
H_{int}=\frac{1}{L^{2}}\sum_{\bm{k}_{1},\dots,\bm{k}_{4}}V_{\bm{k}_{1},\dots,\bm{k}_{4}}t_{\bm{k}_{4}}^{\dagger}t_{\bm{k}_{3}}^{\dagger}t_{\bm{k}_{2}}t_{\bm{k}_{1}}\delta_{\bm{k}_{1}+\bm{k}_{2}-\bm{k}_{3}-\bm{k}_{4}},
\end{equation}
where $L$ is the linear dimension of the system. The details of $V_{\bm{k}_{1},\dots,\bm{k}_{4}}$
do not matter, as long as it is real (thus preserving $C_2\mathcal{T}_+$ symmetry)
and smoothly decays over a distance $\delta k_{t}$ in momentum space (i.e.,
its range is $a_{t}\sim1/\delta k_{t}$ in real space). $V_{\bm{k}_{1},\dots,\bm{k}_{4}}$
can be computed, in principle, by calculating the scattering amplitude
of a pair of trion excitations from momenta $\left(\bm{k}_{1},\bm{k}_{2}\right)$
to $\left(\bm{k}_{3},\bm{k}_{4}\right)$. Three-trion interactions
and higher order many-body terms are also allowed, but become successively
unimportant in the dilute trion limit, as we shall now discuss. 

The effective Hamiltonian (\ref{eq:Heff_SI}) is tractable in the limit
$k_{t}a_{t}\ll1$. Physically, in this limit, the typical distance
between trions present in the ground state, which is of the order
of $1/k_{t}$, is much larger than the size of a trion, $a_{t}$.
In this limit, the effects of the inter-trion interaction are expected
to be small. Indeed, expanding $V_{\bm{k}_{1},\bm{k}_{2},\bm{k}_{3},\bm{k}_{4}}$
at small momenta and using the property $V_{\bm{k}_{1},\bm{k}_{2},\bm{k}_{3},\bm{k}_{4}}=-V_{\bm{k}_{1},\bm{k}_{2},\bm{k}_{4},\bm{k}_{3}}=-V_{\bm{k}_{2},\bm{k}_{1},\bm{k}_{3},\bm{k}_{4}}$
(which follows from the Pauli principle) and the rotational symmetry
of the effective Hamiltonian, we obtain
\begin{equation}
V_{\bm{k}_{1},\bm{k}_{2},\bm{k}_{3},\bm{k}_{4}}=V_{0}a_{t}^{2}(\bm{k}_{1}-\bm{k}_{2})\cdot(\bm{k}_{3}-\bm{k}_{4})+O(k^{4}).
\end{equation}
The absence of a constant term in the limit $k\to 0$ is characteristic of spinless fermions, which cannot occupy the same point in space. 

Since the typical momentum of trions in the ground state is $k_{t}$,
the interaction matrix elements are suppressed by a factor $(k_{t}a_{t})^{2}\ll1$.
This justifies a perturbative treatment of $H_{int}$. To zeroth order,
$H_{{\rm eff}}$ is non-interacting; it is identical to the three-band
Hamiltonian (Eq. 4 of the main text), where the trions are treated as a new `flavor' of
free fermions. The resulting zeroth-order spectrum is gapped, with
a gap $E_{g}\sim |\Delta| k_t$. Turning on infinitesimal interactions
renormalizes the magnitude of the gap, but does not close it immediately. 

To verify the consistency of the perturbative analysis, we estimate
the first-order interaction correction to the energy of a trion excitation,
$\delta E$, and compare it to the gap. Using perturbation theory,
we obtain $\delta E\sim V_{0}\left(a_{t}k_{t}\right)^{2}n_{t}\sim m_{t}V_{0}\left(a_{t}k_{t}\right)^{2}\cdot\varepsilon_{t,0}$, 
where $n_{t}\sim k_{t}^{2}$ is the density of trions in the ground
state. With our choice $\Delta \sim \varepsilon_{t,0}/k_t$, we find that the zeroth order gap satisfies $E_{g}\sim \varepsilon_{t,0} \gg \delta E$. Hence, we have obtained a fully gapped state. 

Finally, the effective Hamiltonian (\ref{eq:Heff_SI}) is manifestly
$C_2\mathcal{T}_+$ symmetric. Our perturbative solution of $H_{{\rm eff}}$
in the limit $k_{t}a_{t}\ll1$ does not break $C_2\mathcal{T}_+$ (or any other
symmetry) spontaneously, as can be seen from the fact that the resulting
ground state in the limit $k_{t}a_{t}\rightarrow0$ is a $C_2\mathcal{T}_+$
symmetric Slater determinant. Infinitesimal interactions cannot cause
spontaneous symmetry breaking in a gapped state, and hence $C_2\mathcal{T}_+$
is preserved for a finite range of $k_{t}a_{t}$.

\subsection{Model for fragile topological bands with $C_3$ symmetry}
\label{sec:C3}
We briefly review a model for fragile topological bands on a honeycomb lattice, satisfying $C_3$, $C_2 \mathcal{T}_+$, and $M_y$ symmetries~\cite{Zou2018Band,Song2019}. The model contains two orbitals of spinless electrons on each sublattice of the honeycomb lattice, such that there are four bands in total. As before, the construction starts from two degenerate pairs of Chern bands with opposite chiralities, each forming a Haldane model~\cite{Haldane1988}. We then add a term that couples the lower two Chern bands, separating them everywhere in the BZ except at the $K$ and $K'$ points.

The Hamiltonian is given by
\begin{align}
H & =-2t_{1}\left[f_{x}(\bm{k})\sigma^{x}+f_{y}(\bm{k})\sigma^{y}\right]\nonumber \\
 & -2t_{2}\,g(\bm{k})\tau^{z}\sigma^{z}-2t_3\left[f_{x}(\bm{k})\sigma^{x}+f_{y}(\bm{k})\sigma^{y}\right]\tau^{x},
 \label{eq:honey}
\end{align}
where $\sigma^{x,y,z}$ act on the sublattice degrees of freedom, $\tau^{x,y,z}$ act on the two orbitals, $t_1$ is the amplitude for nearest-neighbor hopping that preserves the orbital index, $t_2$ is a next-nearest neighbor, orbital-dependent hopping amplitude, and $t_3$ is a nearest-neighbor orbital-flipping hopping term. 
The functions $f_{x,y}$ and $g$ are defined as:
\begin{align}
f_x(\bm{k}) & = \cos\left(\frac{\sqrt{3}}{2}k_{x}\right)\cos\left(\frac{k_{y}}{2}\right)+\cos(k_{y}),\nonumber \\
f_y(\bm{k}) & = \cos\left(\frac{\sqrt{3}}{2}k_{x}\right)\sin\left(\frac{k_{y}}{2}\right)-\sin(k_{y}),\nonumber \\
g(\bm{k}) & = \sin\left(\frac{\sqrt{3}}{2}k_{x}\right)\left[\cos\left(\frac{\sqrt{3}}{2}k_{x}\right)-\cos\left(\frac{3}{2}k_{y}\right)\right].
\end{align}

\begin{figure}[b]
\centering
\includegraphics[width=.45\textwidth]{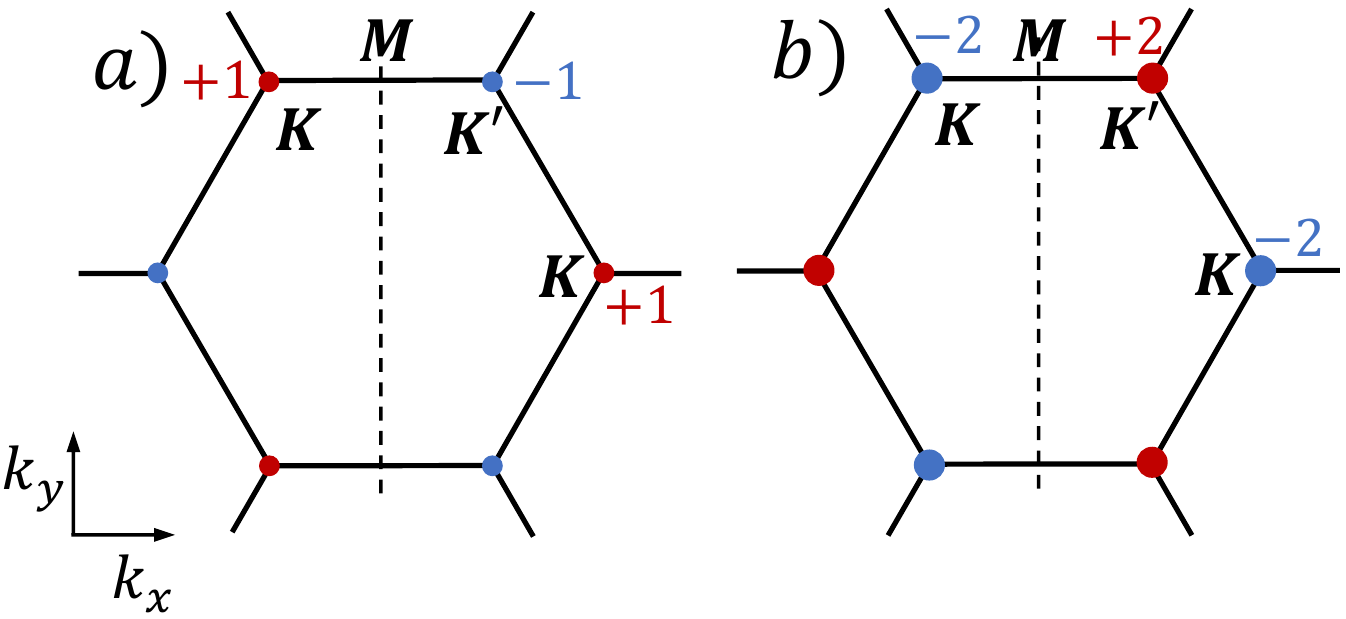}
\caption{(a) Dirac points in the BZ of a graphene-like honeycomb lattice. The two Dirac points at $K$ and $K'$ have opposite winding numbers. (b) After a deformation of the single-particle Hamiltonian, the $K$ and $K'$ points carry quadratic band touchings with opposite winding numbers. The deformation can be done without breaking any lattice symmetry.}
\label{fig:S3}
\end{figure}

The symmetries of the model are represented as:
\begin{align}
C_{2}\mathcal{T}_+ & =\sigma^{x}\tau^{x}K,\nonumber \\
C_{3} & =(\bm{k}\rightarrow R_{3}^{z}\bm{k}),\nonumber \\
M_{y} & =\tau^{x}\times(k_{x}\rightarrow-k_{x}),
\end{align}
where $R_3^z$ denotes a rotation matrix by $2\pi/3$ around the $z$ axis.

Similarly to the square lattice model presented in the main text, the honeycomb model~\eqref{eq:honey} yields two pairs of bands that possess fragile topology. Each pair of bands touch contains two Dirac points at the $K$ and $K'$ points in the BZ. The fragile topological character of the bands can be deduced by computing the winding numbers of the two Dirac points, as in the square lattice model. Alternatively, as explained in Ref.~\cite{Zou2018Band}, due to the mirror symmetry $M_y$ of the model, there is a simple way to determine the relative chirality of the two Dirac points. This is done by examining the $M_y$ eigenvalues of the two eigenstates of the lower bands. Their product can be checked to be $-1$, confirming that the winding numbers of the Dirac points are the same~\cite{Zou2018Band}.

\subsection{Featureless insulator on the honeycomb lattice}

Here, we show that the method we applied in the main text for gapping the Dirac points of fragile topological bands with $C_3$ symmetry can be used to show that a system of electrons on a honeycomb lattice at half filling can form a featureless insulator ground state. 
As in the fragile case, the featureless insulator, that does not break any symmetry and has no topological order, is impossible in the absence of interactions \cite{kimchi2013featureless}. The existence of a featureless insulator on the honeycomb lattice was derived using a different approach in Ref. \cite{Lee2018}. 

As in the case of fragile topological bands, our mechanism proceeds by deforming the single-particle Hamiltonian to form quadratic band touchings, that can then become gapped by hybridizing with a band of trion excitations. We start from a graphene-like single-particle Hamiltonian, in which the spectrum possesses two Dirac points at $K$ and $K'$ (see Fig. \ref{fig:S3}a). In contrast to the fragile case, the two Dirac points carry opposite winding numbers. On the honeycomb lattice, the mirror symmetry ${M}_y$ that takes $k_x \to -k_x$ flips the winding number. This can be seen, e.g., from the fact that the two Bloch states at the $M$ point have the same ${M}_y$ eigenvalue \cite{Zou2018Band}. 

This property of mirror symmetry allows us to smoothly deform this band structure to create quadratic band touchings. To achieve this, we nucleate a pair of new Dirac points with opposite winding numbers at each $M$ point, and shift one of the Dirac points toward $K$ and the other toward $K'$. When this is done in a $C_3$ symmetric way, three Dirac points with winding number $-1$ merge with the original Dirac point of winding $+1$ at the $K$ point, leaving a quadratic band touching of winding $-2$ (see Fig. \ref{fig:S3}b). By mirror symmetry, this procedure leaves the $K'$ point with winding $+2$. Each quadratic band touching can now be gapped following a similar mechanism to that described in the main text, by forming a trion band from states away from the $K$ and $K'$ points and hybridizing them with electrons near those points.

\end{document}